%% file: submission.tex
\documentclass[sigconf, nonacm]{acmart}
\AtBeginDocument{%
  }

\acmSubmissionID{}

\input{sections/header}             % Header content to add before the document (packages/commands/figures)

\begin{document}

%%
%% The "title" command has an optional parameter, allowing the author to define a "short title" to be used in page headers.

% Enhancing LLM-Generated UI with a UI Preference Dataset
% \title{Crowdsourced UI Preference Library-Enhanced UI Generation Powered by LLMs}

\title{Streaming, Fast and Slow: Cognitive Load-Aware Streaming for Efficient LLM Serving}

%%
%% The "author" command and its associated commands are used to define the authors and their affiliations.
%% Of note is the shared affiliation of the first two authors, and the "authornote" and "authornotemark" commands used to denote shared contribution to the research.

\author{Chang Xiao}
% \authornote{Major work completed while Chang Xiao was a Research Scientist at Adobe Research.}
\affiliation{%
  \institution{Boston University}
  \city{Boston}
  \country{USA}}
\email{cxiao1@bu.edu}

\author{Brenda Z. Yang}
\affiliation{%
  \institution{Columbia University}
  \city{New York}
  \country{USA}}
\email{zy2231@columbia.edu}

%%
%% By default, the full list of authors will be used in the page headers. Often, this list is too long, and will overlap other information printed in the page headers. This command allows the author to define a more concise list of authors' names for this purpose.
% \renewcommand{\shortauthors}{Liu \etal}

%%
%% The abstract is a short summary of the work to be presented in the article.
\input{sections/0-abstract}

\begin{CCSXML}
<ccs2012>
   <concept>
       <concept_id>10003120.10003121.10003124.10010870</concept_id>
       <concept_desc>Human-centered computing~Natural language interfaces</concept_desc>
       <concept_significance>500</concept_significance>
       </concept>
 </ccs2012>
\end{CCSXML}

\ccsdesc[500]{Human-centered computing~Natural language interfaces}

%%
%% Keywords. The author(s) should pick words that accurately describe the work being presented. Separate the keywords with commas.
\keywords{Cognitive Load; Large Language Models; Human-AI Interaction; Adaptive Text Streaming
}

\begin{teaserfigure}
  \centering
  \includegraphics[width=\textwidth]{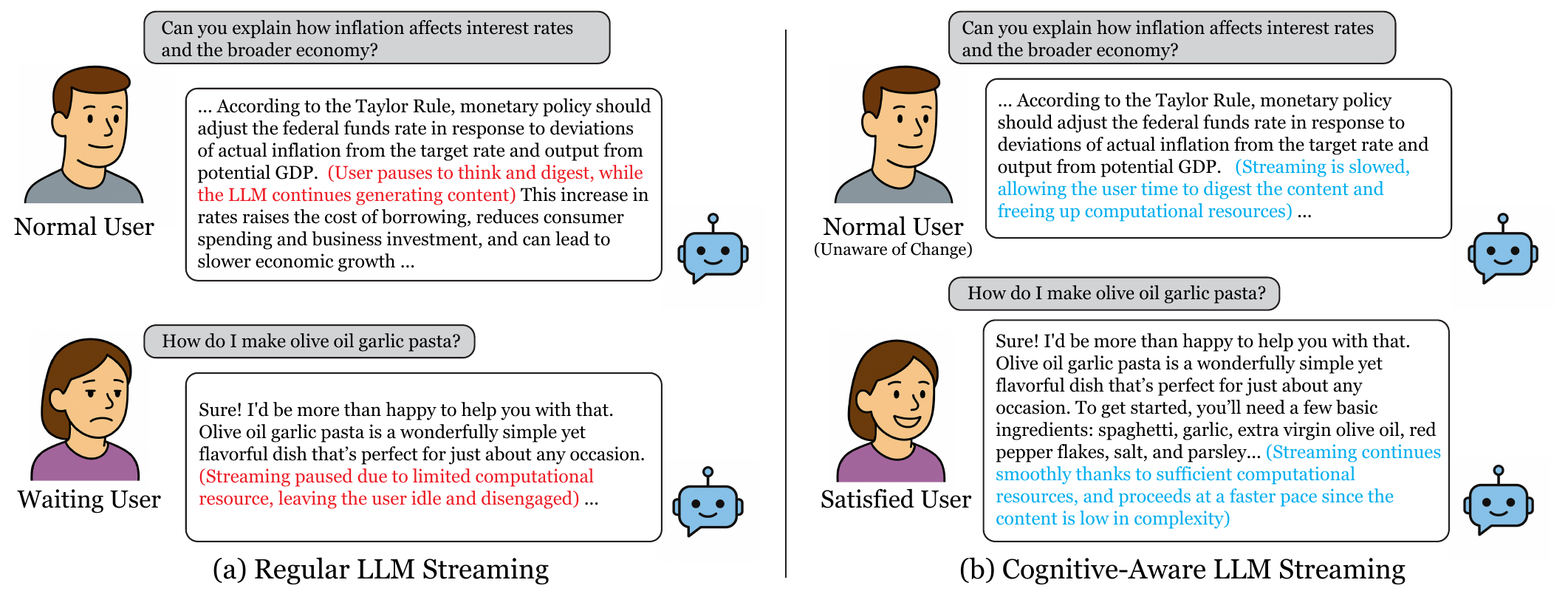}
  \caption{(a) In regular LLM streaming, the streaming speed is not associated with the content being delivered. Complex content may be streamed faster than users can comfortably read, resulting in wasted resource. Conversely, simple content may be streamed too slowly, causing users to wait unnecessarily.
(b) In our proposed approach, the streaming speed is adapted based on the estimated content complexity. Complex content is slowed down to support user comprehension and optimize resource usage, while simpler content is streamed more quickly to better align with the user's natural reading speed.}\label{fig:teaser}
  \Description{A descriptive text for screen readers.}
\end{teaserfigure}

\maketitle

\input{sections/document}       
\bibliographystyle{ACM-Reference-Format}
\bibliography{sections/bibliography}

% \appendix

% \input{supplemental material/appendix}

\end{document}

%% file: sections/header.tex
\input{commands_packages/published-packages} % New (approved) packages

\input{commands_packages/dev-packages} % New packages for custom comments
\input{commands_packages/dev-commands} % Custom commands / styling / comments

%% file: commands_packages/published-packages.tex
\usepackage{xspace}     % Used for abbreviation spacing
\usepackage{xpunctuate} % Used for abbreviation spacing

%% file: commands_packages/dev-packages.tex
\usepackage{xargs}      % Used for new commands with optional arguments
\usepackage{soul}       % Used for custom comments
\usepackage{color}      % Used for custom colors in comments
\usepackage{xcolor}
\usepackage{colortbl}
\usepackage{enumitem}
\usepackage{soul}
\usepackage{subcaption}
\usepackage{multirow}
\usepackage{circledsteps}
\usepackage{hyperref}
\usepackage{algorithm}
\usepackage{algorithmic}

\usepackage{listings}
\lstdefinelanguage{json}{
    basicstyle=\tiny\ttfamily,
    stepnumber=1,
    numbersep=8pt,
    showstringspaces=false,
    breaklines=true,
    frame=none,
    literate=
     *{0}{{{\color{blue}0}}}{1}
      {1}{{{\color{blue}1}}}{1}
      {2}{{{\color{blue}2}}}{1}
      {3}{{{\color{blue}3}}}{1}
      {4}{{{\color{blue}4}}}{1}
      {5}{{{\color{blue}5}}}{1}
      {6}{{{\color{blue}6}}}{1}
      {7}{{{\color{blue}7}}}{1}
      {8}{{{\color{blue}8}}}{1}
      {9}{{{\color{blue}9}}}{1}
      {:}{{{\color{black}:}}}{1}
      {,}{{{\color{black},}}}{1}
      {\{}{{{\color{orange}\{}}}{1}
      {\}}{{{\color{orange}\}}}}{1}
      {[}{{{\color{orange}[}}}{1}
      {]}{{{\color{orange}]}}}{1},
}
\definecolor{darkgreen}{rgb}{0.0, 0.5, 0.0}  % RGB value for dark green
\lstdefinelanguage{markdown}{
    basicstyle=\tiny\ttfamily,
    morekeywords={\#, \*\*, \*, -}, % Escape special characters
    keywordstyle=\color{blue},
    sensitive=false,
    morecomment=[l]{>}, % Define blockquotes
    commentstyle=\color{gray}\itshape,
    morestring=[b]"
}

%% file: commands_packages/dev-commands.tex
\definecolor{LIGHTPINK}{RGB}{237,157,202}
\definecolor{LIGHTRED}{RGB}{210,121,121}
\definecolor{LIGHTORANGE}{RGB}{230,170,50}
\definecolor{LIGHTGOLD}{RGB}{210,194,121}
\definecolor{LIGHTGREEN}{RGB}{121,210,121}
\definecolor{LIGHTAQUA}{RGB}{121,206,210}
\definecolor{LIGHTBLUE}{RGB}{121,124,210}
\definecolor{LIGHTPURPLE}{RGB}{153,102,255}
\definecolor{RED}{RGB}{178,34,34}
\definecolor{GRAY}{RGB}{166,166,166}
\definecolor{WHITE}{RGB}{255,255,255}

\newcommandx{\yimeng}[2][1=] 
    {\setulcolor{LIGHTGREEN}{\ul{#1}} \textcolor{LIGHTGREEN}
    {[\textbf{Yimeng:} #2]}}
\newcommandx{\guest}[3][1=]
    {\setulcolor{LIGHTORANGE}{\ul{#1}} \textcolor{LIGHTORANGE} 
    {[\textbf{#2:} #3]}}

\newcommand\circled[2][]{
\ifmmode
    \Circled[fill color=black,inner color=white, inner ysep=5pt, inner ysep=5pt, #1]{\mathsf{#2}}
    \else
    \Circled[fill color=white,inner color=black, inner ysep=5pt, inner ysep=5pt, #1]{\sffamily#2}
\fi
}

%% file: sections/0-abstract.tex
\begin{abstract}

Generative conversational interfaces powered by large language models (LLMs) typically stream output token-by-token at a rate determined by computational budget, often neglecting actual human reading speeds and the cognitive load associated with the content. This mismatch frequently leads to inefficient use of computational resources. For example, in cloud-based services, streaming content faster than users can read appears unnecessary, resulting in wasted computational resources and potential delays for other users, particularly during peak usage periods. To address this issue, we propose an adaptive streaming method that dynamically adjusts the pacing of LLM streaming output in real-time based on inferred cognitive load. Our approach estimates the cognitive load associated with streaming content and strategically slows down the stream during complex or information-rich segments, thereby freeing computational resources for other users. We conducted a statistical analysis and simulation based on a statistical model derived from data collected in a crowdsourced user study across various types of LLM-generated content. Our results show that this adaptive method can effectively reduce computational consumption while largely maintaining streaming speed above user's normal reading speed.

\end{abstract}

%% file: sections/document.tex
%!TEX root = proceedings.tex

\input{sections/1-introduction}

\input{sections/2-relatedwork}

\input{sections/2.5-analysis}
\input{sections/3-system}

\input{sections/4-evaluation}
\input{sections/5-conclusion}

%% file: sections/1-introduction.tex
\section{Introduction}

Conversational AI have rapidly gained popularity and utility since their release. Trained on large-scale corpora using high-capacity deep models, these systems offer highly interactive and informative experiences, allowing users to access knowledge and engage in conversations across diverse domains simply by asking questions in natural language~\cite{OpenAI2023}.

Most user-facing LLM services, including widely used platforms like ChatGPT, are hosted on cloud infrastructures. While this setup offers broad accessibility, it also poses significant challenges in computational resource management. 
As of early 2025, ChatGPT has experienced substantial growth in user engagement, serving over 400 million weekly active users who collectively send more than 1 billion messages daily~\cite{ChatGPTStats2025}. During peak usage periods, demand frequently exceeds the available computational capacity, causing user requests to queue, leading to increased waiting times and degraded user experiences~\cite{ChatGPTOutage2025}. Figure \ref{fig:teaser}(a) illustrates a common scenario: when a user submits a query, the LLM generates output at its maximum available speed. However, rapid generation may be unnecessary when the content is dense or cognitively demanding---for example, complex sentence structures, intricate logical arguments, or uncommon terminology---as users often need to pause or slow down their reading.  Despite this, the LLM continues streaming at full pace regardless of content, resulting in wasted computational resources. At the same time, other users may experience prolonged waiting times during peak periods due to insufficient resources. Thus, there is a growing need for adaptive optimization of text streaming in LLMs to enhance both system efficiency and user experience.

Prior to the emergence of LLMs, adaptive text streaming had received little attention in both research and practice. This was largely because traditional text-based content (\textit{e.g.,} articles, emails, or web pages) was typically delivered either in full or at speeds far exceeding human reading rates, allowing users to consume the content at their own pace. As a result, there was minimal concern about computational resource constraints or the need for fine-grained control over text delivery speed.

% However, LLM requires substantial computational power, and can only generate content in a token-by-token manner and stream it to users

% [LLM cannot show text instantly, as it computed token by token, and takes time to compute the next token. It naturally enforce the user consume content is a sequential way. Also, LLM generate speed is content irrelevant.]
In contrast, modern chat-based LLMs have brought the challenge of text streaming to the forefront. Because LLMs require substantial computational power and generate content token by token, their responses are typically streamed to users at a rate determined by the available computational resources and model complexity. This means that users must consume the content incrementally, as it is generated in real time. 
% Notably, unlike website content or email, which deliver text almost instantly, the generation speed of online LLM services is on a similar order of magnitude to the average human reading speed. For example, popular online LLM chatbots like ChatGPT or Deepseek generate text at approximately 500-5,000 words per minute (WPM)~\cite{ArtificialAnalysis2025GPT4o}, depending on the model used and the underlying server infrastructure, whereas typical adult reading speeds range from about 100 WPM (technical documents) to 700 WPM (skimming and scanning)~\cite{carver1990reading}.
Although in some scenarios users may wait for the full output to be generated before reading (e.g., in code generation), in practice, most chat-based LLM interfaces are designed to support immediate consumption as whenever output tokens becomes available. This streaming interaction model has become the norm because it gives users quicker feedback, a greater sense of responsiveness, and the opportunity to interrupt or re-prompt based on partial output~\cite{chutani2024streaming, liu2024andes, li2025adaserve, li2024eloquent}.
Crucially, this shift transforms text from a static payload into a dynamic, time-based medium. Combined with the cloud-hosted nature of LLMs and their massive user bases, it raises timely and important questions about how to stream text efficiently and adaptively for better computational resource management.

\begin{figure}[t]
  \centering
  \includegraphics[width=1.0\linewidth]{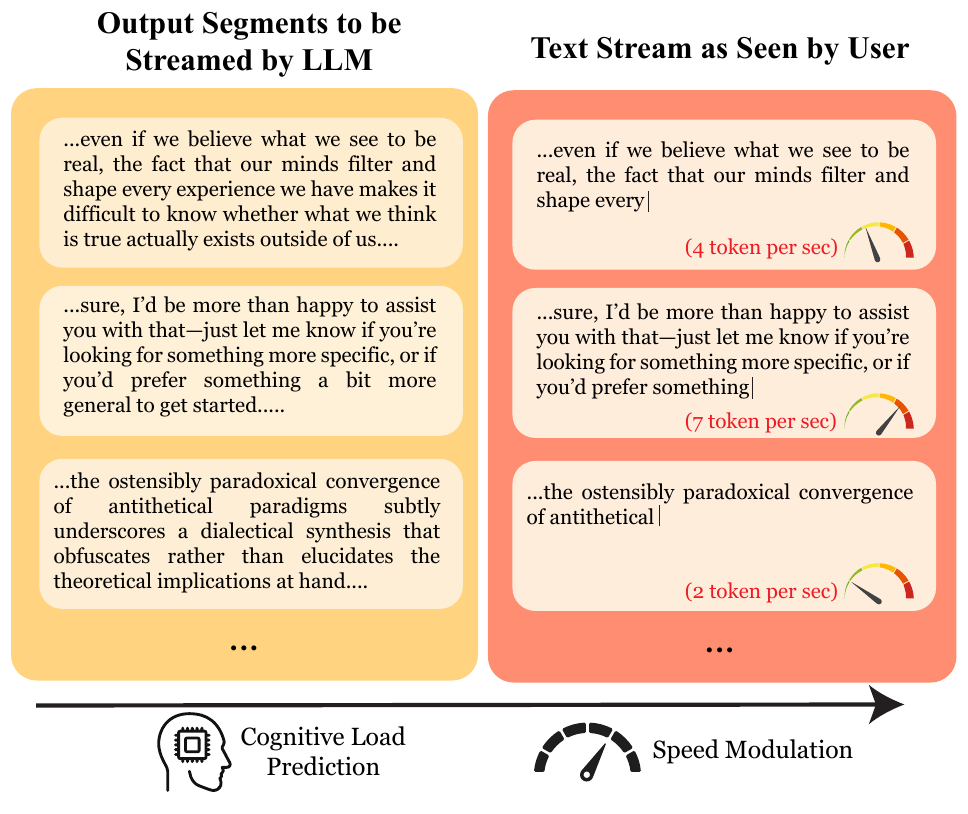}
  \caption{Overview of the workflow of our cognitive-aware streaming framework. Content with high cognitive load is streamed at a lower speed to give users more time to process it, while content with low cognitive load is streamed at a higher speed to minimize unnecessary waiting.}
  \label{fig:overview}
\end{figure}

Traditional adaptive streaming methods employed in multimedia services typically rely on static or rule-based strategies, such as buffer-based throttling~\cite{huang2013downton} or quality adjustment mechanisms~\cite{rejaie2000layered}. While these techniques could be adapted for LLM-based services, they primarily operate from a system-level perspective and overlook the significant differences in users' cognitive load when reading various LLM-generated content. Variations in complexity and informational density can directly affect the reading speed of users. This situation poses an intriguing research question: \textbf{Can we dynamically identify text segments where users experience increased cognitive load and strategically adjust the streaming pace, thereby reallocating computational resources more efficiently?}

To address this, we propose a novel adaptive streaming framework specifically designed for LLM outputs by incorporating real-time cognitive load assessment to modulate streaming speeds. Figure \ref{fig:overview} provides a high-level overview of the system architecture.
% [this study focus on application requires users attention, no code generation]
We conducted a statistical analysis and simulation based on a statistical model derived from data collected in a crowdsourced user study ($N=300$) across various types of LLM-generated content. The results show that, compared to a standard streaming strategy, our approach can significantly reduce computational resource usage while still delivering content above the user’s normal reading speed.

In summary, the contributions of this paper are threefold:

\begin{itemize}
    \item We introduce the concept of adaptive text streaming based on the cognitive load of the content to be streamed, aimed at improving the efficiency of LLM serving in conversational AI applications.
    
    \item We present a statistical analysis framework that enables practitioners to quantify potential computational resource savings in their own scenarios.
    
    \item We develop a prototype system that estimates user reading speed and dynamically adjusts the streaming speed. 
    Our experiments demonstrate that our method effectively reduces computational usage while largely ensuring that users receive content above their normal reading speed.
\end{itemize}

%% file: sections/2-relatedwork.tex
%!TEX root = proceedings.tex

%%
%% NOTE: Assign collaboration badges and section labels to all sections and 
%% subsects when created (badges include: \incomplete, \underRevision,
%% \readyForFeedback, \feedbackProvided, \complete, \locked)
%%
\section{Related Work}
\label{sec:relatedwork}

\label{sec:system}

\subsection{Human Reading Speed of Textual Content}\label{sec:cog_assess}

Reading speed in digital environments is a topic of active research across HCI, cognitive psychology, education, and information science~\cite{kurniawan2001reading, franken2015eye}. Numerous factors --- from display technology~\cite{siegenthaler2012lcd} and screen size~\cite{elliott2020effect} to typography~\cite{wallace2022towards}, layout~\cite{dyson2001influence}, and user interface design~\cite{oquist2007eye} --- have been studied for their impact on how quickly people can read text on screens. 

While these factors play an important role in shaping how humans consume text, they typically remain fixed during a user’s interaction with a conversational AI system.
Instead, what varies is the content itself, as users submit different requests and receive diverse responses. Therefore, the textual content itself---and more specifically, the cognitive load it imposes---becomes the dominant factor influencing how quickly a user can process the information~\cite{destefano2007cognitive, fink2001speed}. Consequently, cognitive load can be used as the primary signal for estimating human reading speed in the context of human-LLM interactions.

\paragraph{Cognitive Load Estimation}

The Cognitive Load Theory (CLT)~\cite{plass2010cognitive, sweller1988cognitive}, one of the most widely accepted frameworks for understanding how humans process information, categorizes cognitive load into three types: intrinsic load, determined by the complexity of the content (e.g., how much information the content contains); extraneous load, related to how the information is presented (e.g., the complexity of sentence structures); and germane load, referring to the reader’s active effort in integrating the information (e.g., mentally summarizing content or making inferences).

Due to the internal nature of cognitive processes, directly measuring cognitive load is challenging. Various studies have attempted to infer it using physiological and behavioral signals collected during reading. Eye-tracking metrics (e.g., fixation duration, regressions, pupil dilation)~\cite{joseph2020potential, zagermann2016measuring} and EEG signals (e.g., increased theta and reduced alpha activity)~\cite{kosch2023survey, kumar2016measurement} have been shown to correlate strongly with mental effort. Additional indicators such as heart rate variability~\cite{solhjoo2019heart} and electrodermal activity~\cite{buchwald2019electrodermal} also reflect sustained cognitive strain. These signals provide real-time, user-specific feedback and have been used in multimodal datasets like ZuCo~\cite{hollenstein2018zuco} and CLARE~\cite{bhatti2024clare} to train models for cognitive load prediction. 

While these methods offer accurate measurement of cognitive load, they require specialized hardware and calibration, which limits their practicality in the context of human-LLM interaction. Therefore, we focus on metrics that can be directly calculated from the text content itself to infer cognitive load.

\paragraph{Text-Based Cognitive Load/Readability Estimation} 

Readability refers to the ease with which readers can comprehend written text and has been shown to correlate strongly with the cognitive load experienced by humans when processing textual content~\cite{crossley2019moving}. 
% In this work, we use readability and cognitive load interchangeably to describe the cognitive demand required to read and understand a given piece of text.
Traditionally, readability has been approximated using surface-level linguistic features such as sentence length, word length, and vocabulary frequency~\cite{dale1949concept}.
Classic text difficulty formulas, including Flesch-Kincaid~\cite{kincaid1975derivation}, Dale-Chall~\cite{dale1948formula}, and Gunning Fog~\cite{gunning1952technique}, provide fast estimates of readability but often fail to capture deeper semantic, syntactic, or discourse-related complexity~\cite{crossley2023large}.
To address these limitations, researchers have proposed syntactic complexity metrics~\cite{basili1983empirical} (e.g., clause density, parse tree depth) and cohesion indices (e.g., referential overlap, connectives) to more accurately reflect the processing effort required for comprehension.
Tools such as Coh-Metrix~\cite{graesser2004coh} compute hundreds of such features to model cognitive load more comprehensively. While these methods are interpretable and efficient, they are limited by their reliance on static textual features and often fail to account for longer-range contextual factors.

Machine learning approaches have advanced the estimation of cognitive load during reading by learning directly from annotated datasets rather than relying solely on predefined rules. Early models combined linguistic features with classifiers or regressors to predict reading level or cognitive demand~\cite{franccois2012nlp, pilan2014rule}.
More recent work leverages pre-trained transformer-based language models such as BERT~\cite{devlin2019bert} and GPT~\cite{radford2019language}, fine-tuned on datasets including WeeBit~\cite{vajjala2012weebit}, OneStopEnglish~\cite{vajjala2018onestopenglish}, and Newsela~\cite{xu2015problems}. These models outperform traditional approaches by capturing rich syntactic and semantic patterns, although they often lack interpretability and require substantial labeled data.
Despite these trade-offs, ML-based approaches offer scalable and accurate cognitive load estimation, making them well-suited for real-time applications. Recent studies further show that LLMs can provide general-purpose assessments of cognitive demands in longer and more complex paragraphs~\cite{liu2025automatic, patel2023improving}. By leveraging world knowledge and simulating human-like comprehension, LLMs offer a more integrated and context-aware framework for estimating cognitive load from text.

% In the context of human-LLM interaction, subjective reports and physiological signals are difficult to collect. 

% Given that most LLM interactions occur through text modality, reading time emerges as the most practical indicator for estimating users' cognitive load, which is also supported by previous studies~\cite{cirilo1980text, raney1993monitoring}.

% This section surveys general readability assessment methods, as readability estimation forms a key component of our proposed framework.

% \paragraph{Readability Measurement}

% Readability refers to the ease with which readers can comprehend written text and has been shown to correlate highly with the cognitive load experienced by humans when processing textual content~\cite{crossley2019moving}.

\subsection{Resource Optimization for Streaming Output}
\paragraph{Efficient LLM Serving}
As cloud-based LLMs like ChatGPT scale to serve over a billion user interactions daily, the efficient delivery of AI-generated content has become a critical concern. Recent work has proposed system-level optimizations that treat streaming timing as a tunable parameter to improve both latency and resource utilization. For instance, Andes~\cite{liu2024andes} introduces a Quality-of-Experience-aware scheduling system that reallocates GPU time by detecting when users accumulate unread tokens, temporarily pausing streams to better balance serving capacity across users. Similarly, AdaServe~\cite{li2025adaserve} leverages fine-grained speculative decoding to dynamically adjust token generation speed to improve serving efficiency. TimelyLLM~\cite{ling2024timelyllm} focuses on robotic applications of LLMs, efficiently optimizing the LLM serving schedule by leveraging the time redundancy between robot plan generation and execution.
On the transmission layer, Eloquent~\cite{li2024eloquent} improves token delivery over the network by merging newly generated tokens and currently unacknowledged tokens in the next outgoing packet~\cite{wallace2022towards}. 

However, these methods focus primarily on system-level metrics and overlook the variability in LLM-generated content, which often differs in readability and affects how quickly users can consume the output.

\paragraph{Adaptive Text Delivery}
Beyond LLM use cases, researchers have explored adaptive text delivery in other reading interfaces and intelligent tutoring systems. Early work like Time Aura~\cite{mamykina2001time} introduced ambient visual pacing interfaces that helped users regulate their pace during cognitively demanding tasks such as test-taking or presentations. Lander et al.~\cite{lander2015collaborative} developed gaze-adaptive scrolling that uses eye-tracking to adjust the pace of text scrolling for public display. In immersive environments, Grootjen et al.~\cite{grootjen2024your} presented a VR system that adapts word speed based on pupil dilation, a proxy for cognitive load, to slow down when users are overwhelmed and speeding up when they are not. Intelligent Tutoring Systems like AutoTutor~\cite{graesser2004autotutor} and Gaze Tutor~\cite{d2012gaze} similarly adjust the pacing of explanations or interventions based on learner engagement, gaze, or confusion. 

Another extensively explored technique for adaptive text delivery is Rapid Serial Visual Presentation (RSVP), where individual words are displayed sequentially at a fixed screen location. Early pioneering work by Öquist et al.\cite{oquist2001adaptive,oquist2003towards} introduced adaptive RSVP algorithms that modulate presentation timing based on textual features, achieving reduced task load compared to static RSVP. More recently, Kosch et al.\cite{kosch2020one} incorporated EEG-based cognitive load monitoring and found significant correlations between physiological signals and presentation speed, proposing dynamically adjusted RSVP rates based on EEG signals.

\paragraph{Other Scenarios} Adaptive streaming techniques have also been applied in multimodal interfaces other than text. For example, in multimedia classroom, segmentation of instructional videos and adaptive pausing based on gaze or confusion signals has been shown to improve comprehension and retention~\cite{lalle2015prediction, hutt2019time, hutt2021breaking}. In smartphone use, attention-aware systems like Attelia~\cite{okoshi2015attelia} detect cognitive breakpoints in smartphone usage and delay notifications until moments of low cognitive load, reducing disruption and information overload. In narrative visualizations, gaze-driven systems dynamically highlight relevant content as users read, enhancing the coupling between text and data~\cite{barral2020understanding, lalle2019gaze}. In VR, human perception of color has been utilized to adjust the streaming display quality to reduce the power consumption~\cite{duinkharjav2022color}.

These examples show how user attention and perceptual constraints can guide adaptive pacing. The principles behind these systems---detecting cognitive state and dynamically adjusting delivery pace---can inform the design of text streaming systems in LLMs.

%% file: sections/2.5-analysis.tex
\section{What is the Benefit of Adaptive Streaming?}\label{sec:analysis}
We begin by defining the scope of our study. First, we focus on scenarios involving human–LLM interaction, where users actively consume the generated content as it is streamed. Other applications of LLM, such as code generation or agent–agent communication, fall outside the scope of this work.

Second, we consider a cloud-based LLM serving environment, where computational resources are shared among multiple users simultaneously. We do not aim to optimize the experience of single-user scenarios, such as locally deployed LLMs.

Last, we adopt a statistical perspective to model and analyze the problem. Our objective is to optimize the aggregated user experience under a given computational budget, rather than tailoring the streaming speed to specific users.

With this scope in mind, the central research question we aim to answer is: \textbf{How much computing resource can be saved by adaptively modulating text streaming speed?} Namely, under what conditions can adaptive streaming improve computational efficiency, and to what extent would it affect overall user experience? Answering these questions helps justify the need for our system. In this section, we develop a statistical framework to explore and quantify these aspects.

\subsection{Quantifying Computing Resource Savings}

Intuitively, content with low information density, simple sentence structures, and familiar vocabulary is easier to read and thus requires a faster streaming speed---streaming too slowly in these disrupts the user's reading flow and degrades their experience. More complex and information-dense content naturally slows down users’ reading pace, allowing for slower streaming without negative impact. Moreover, streaming content faster than a user’s reading speed does not enhance the experience either~\cite{acklin2017modern}, since users will naturally read at their own pace regardless of how quickly the text becomes available. This means any excess delivery speed is effectively unused and instead leads to inefficient use of system resources. This asymmetry motivates an adaptive streaming approach that aligns output speed with the user’s expected reading pace.

%rephrased above paragraph for more accurate wording, and added citation to support claim

\paragraph{Modeling Reading Speed.}
Without loss of generality, we consider a setting in which two groups of users interact with a cloud-based LLM service, where each group consumes generated content with distinct cognitive demands---arising from differences in topic, writing style, or text complexity (e.g., Group A reads narrative stories, while Group B reads technical papers).

Although reading speed varies across individuals due to factors such as personal habits, familiarity with the content, cognitive processing ability, and language proficiency, prior research suggests that users' dwell time on online articles follows a log-normal distribution due to its non-negative and right-skewed nature~\cite{yin2013silence}. Therefore, for each group, we model the user's natural reading speed as a log-normal distribution,

\begin{align*}
    r_A &\sim \text{Lognormal}(\mu_A, \sigma_A^2), \\
    r_B &\sim \text{Lognormal}(\mu_B, \sigma_B^2),
\end{align*}

where $\mu_A$ and $\mu_B$ are the means of the underlying normal distributions (i.e., $\log r$), and $\sigma_A, \sigma_B$ are the corresponding standard deviations. Here, $r$ represents the natural reading speed when the entire passage is fully presented (i.e., not streamed).

%revised assumption
Our core assumption is that the streaming speed should be at least equal to a user’s natural reading speed to prevent noticeable delays in content delivery. That is, for a given group, if we stream at speed $s$ such that

\[
\Pr(r \leq s) \geq \alpha,
\]

then a fraction $\alpha$ of users will receive content at a pace that matches or exceeds their reading speed, while the rest may notice latency due to slower-than-expected delivery. This defines the \textit{Streaming-Reading Alignment Rate }(SRAR), denoted as $\alpha$, which represents the fraction of users whose natural reading speed is less than or equal to the chosen streaming speed $s$.

\textbf{Our goal here is to save computational resources while maintaining a high target SRAR across all users.}

%revise to SRAR, not user satisfaction rate

\paragraph{Savings When Resources Are Sufficient.}

\begin{figure}[t]
  \centering
  \includegraphics[width=1.0\linewidth]{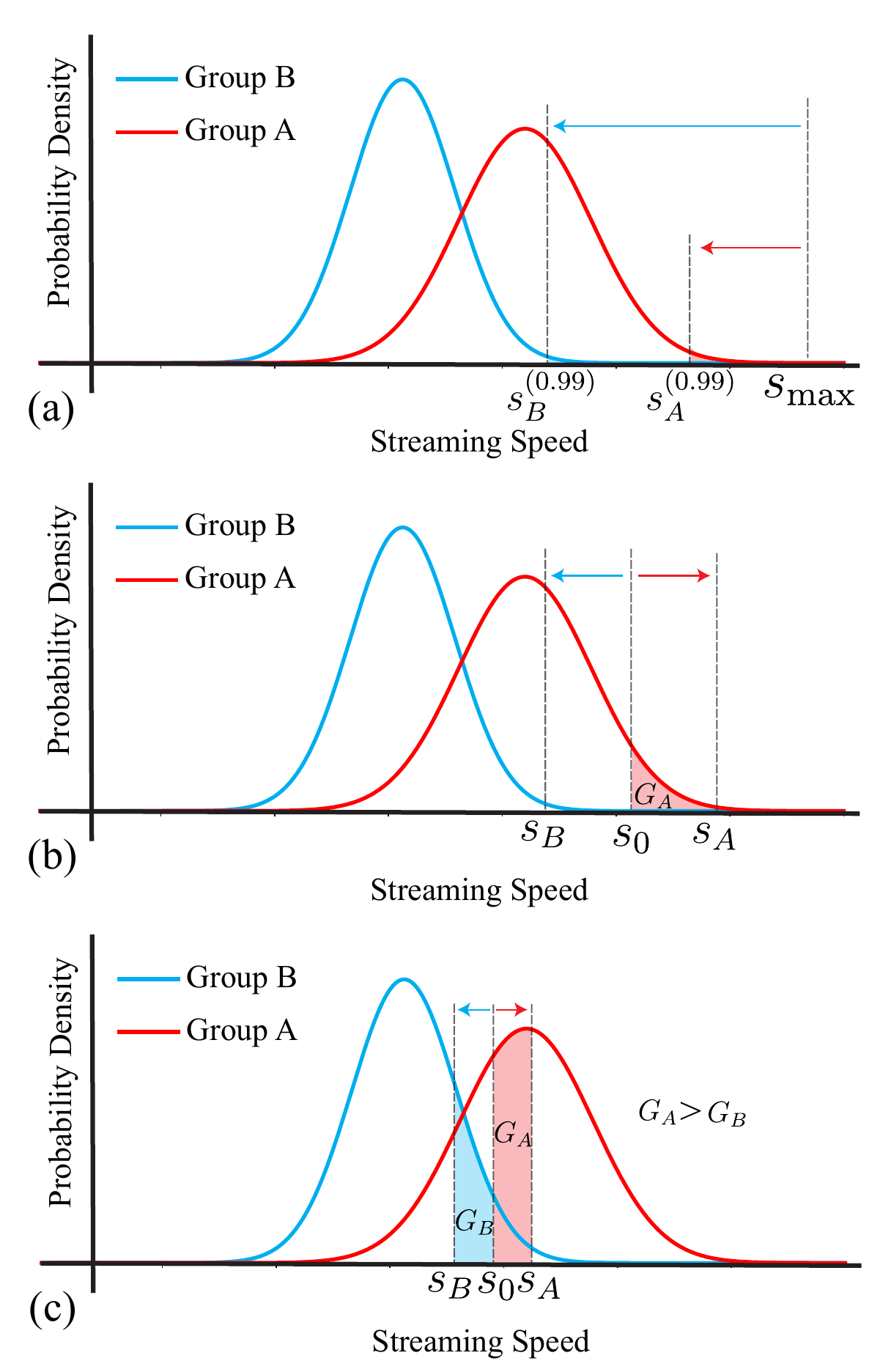}
  \caption{Potential computing resource savings of adaptive streaming under different scenarios. Arrows indicate streaming speed adjustments. (a) When the maximum LLM streaming speed $S_{\text{max}}$ is very high, it can be reduced to $s_A^{(0.99)}$ and $s_B^{(0.99)}$ for Groups A (easier content) and B (harder content), respectively. This maintains a comfortable reading speed for 99\% of users in each group while significantly reducing resource usage. (b) When the available resource can only support both Groups A and B at a streaming speed of $s_0$, which is lower than the thresholds in (a), it is still possible to reduce Group B’s speed to $s_B$ and reallocate the saved resource to increase Group A’s speed. This enhances Streaming-Reading Alignment Rate (SRAR) for Group A (area $G_A$) with minimal impact on Group B’s experience. (c) If the available computing resources is even more constrained, a trade-off can still possibly improve overall SRAR. By reducing Group B's speed and increasing Group A's speed, overall SRAR improves as long as $G_A > G_B$, resulting in a net gain of $G_A - G_B$.}
    \vspace{-3mm}
  \label{fig:gaussian}
\end{figure}

Given the log-normal distribution of reading speeds, the $\alpha$-quantile (i.e., the minimum speed that covers $\alpha$ percent of users) can be computed using the inverse Cumulative Distribution Function of the log-normal distribution:

\begin{align*}
    s_A^{(\alpha)} &= \exp\left( \mu_A + z_\alpha \cdot \sigma_A \right), \\
    s_B^{(\alpha)} &= \exp\left( \mu_B + z_\alpha \cdot \sigma_B \right),
\end{align*}

where $z_\alpha $ is the z-score of a normal distribution corresponding to percentile $\alpha$ (e.g., $z_{0.99} \approx 2.33$, $z_{0.95} \approx 1.64$, and $z_{0.90} \approx 1.28$).

Let $s_{\text{max}}$ denote the maximum streaming speed supported by the system (i.e., streaming at the full generation rate without throttling). In traditional non-adaptive streaming, when resources are sufficient, text is streamed at $s_{\text{max}}$ to all users, regardless of content.

In contrast, our adaptive streaming allows the rate to be reduced to $s_A^{(\alpha)}$ and $s_B^{(\alpha)}$ for each group, while still ensuring a desired quality threshold $\alpha$. Let $p_A$ and $p_B$ be the proportions of users in Groups A and B, respectively, such that $p_A + p_B = 1$. Then the expected computing resource saving under adaptive streaming is

\begin{equation}\label{eq:saving}
    \text{Saving}(\alpha) = 1 - \frac{ p_A s_A^{(\alpha)} + p_B s_B^{(\alpha)}}{s_{\text{max}}}.
\end{equation}

This provides a tunable control knob of a trade-off: higher values of $\alpha$ ensure a better user experience but yield less savings, while lower $\alpha$ values improve efficiency at the cost of allowing more users to experience slower streaming. Figure~\ref{fig:gaussian}(a) illustrates the potential resource savings under this scheme. Suppose we aim to stream at a rate that exceeds the reading speed of 99\% of users in each group. In that case, we can set the streaming speeds to $s_A^{(0.99)}$ and $s_B^{(0.99)}$ for Groups A and B, respectively. The corresponding resource savings can then be computed using Equation~\ref{eq:saving}.

\paragraph{Savings When Resources Are Limited}

We have characterized the potential computing resource savings when the system's maximum streaming speed is sufficiently high. However, in practice, the LLM server may not have enough capacity to stream at those speeds. Suppose the available resources only permit an averaged streaming rate of $s_0$ for both groups, and assume equal group sizes, i.e., $p_A = p_B = 0.5$.

In this constrained scenario, shown in Figure~\ref{fig:gaussian}(b), a non-adaptive strategy would be to stream at $s_0$ uniformly, which will result in a certain number of user-observed delays. Instead, we can redistribute resource by slightly decreasing Group B’s speed to $s_B$, which still maintains a high SRAR in that group, and reallocate the saved resource to increase Group A’s speed to $s_A$. This adjustment enables a new subset of Group A users---denoted as area $G_A$---to receive content at or above their natural reading speed, improving their experience without largely degrading the experience for Group B.

In an even more resource-limited case, illustrated in Figure~\ref{fig:gaussian}(c), the shared rate $s_0$ is further reduced. Here, lowering Group B’s speed to $s_B$ will result in more users in Group B falling below their comfortable reading threshold (represented as area $G_B$), but the reallocated resource allows a larger number of Group A users (area $G_A$) to be covered. If the goal is to maximize overall user SRAR when average speed is limited at $s_0$, this redistribution is beneficial when $G_A > G_B$. If $G_A < G_B$, such move would reduce the overall SRAR and should be avoided.

Overall, when resources are limited, adaptive reallocation of streaming speed can still improve user experience, but must be guided by the marginal gains and losses across groups.

\subsection{A Real Example from Crowdsourcing User Study}\label{sec:real}

\begin{figure}[t]
  \centering
  \includegraphics[width=1.0\linewidth]{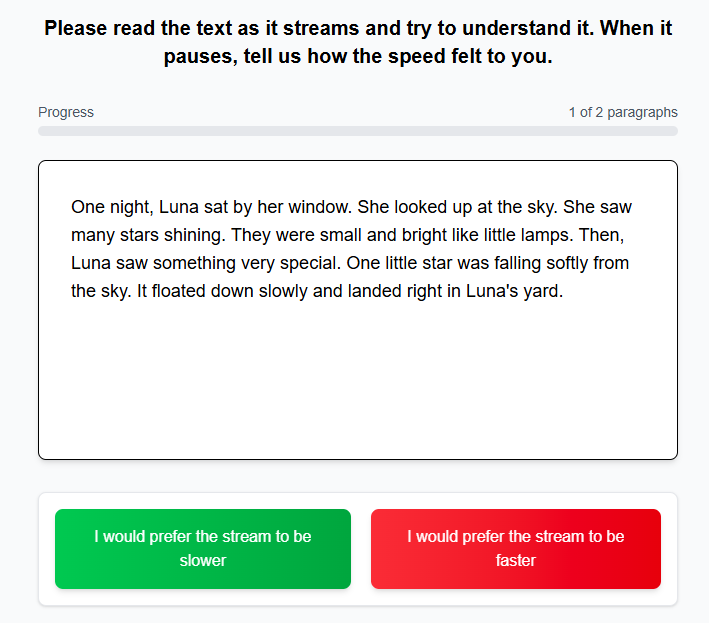}
  \caption{A screenshot of our crowdsourcing web interface.}
    \vspace{-3mm}
  \label{fig:screenshot}
\end{figure}

The previous section introduced a theoretical framework for analyzing potential computing resource savings and balancing the trade-off between efficiency and user experience. Here, we present a real-world text streaming scenario through a crowdsourced user study, offering concrete empirical results. This not only helps validate key theoretical assumptions but also provides practical examples for those looking to apply a similar adaptive streaming approach in their own LLM systems.

\paragraph{Study Protocol}
%%revised: add readability level detail
In this test, we conducted a user study based on two distinct content scenarios: (1) a bedtime story suitable for a six-year-old child, and (2) an explanatory passage on an economic concept. These texts were generated offline using GPT-4o to simulate content streamed by LLM service. We used the Flesch–Kincaid readability tests~\cite{kincaid1975derivation} to verify that the readability level of the bedtime story is below 5th grade, while the economic passage is at the college level. We also explicitly prompt GPT-4o to generate content that maintains sentence-level consistency within each passage.

We recruited 100 participants through the Prolific platform, as prior research suggests that Prolific provides higher-quality data than other crowdsourcing platforms, and sometimes even better than in-person studies~\cite{douglas2023data}. Participants ranged in age from 21 to 68 years (avg. 34.7), with 57 male and 43 female. Additionally, 58\% of participants held a bachelor’s degree or higher. All participants were required to be native English speakers to eliminate the influence of language proficiency. Each participant was asked to read two passages---one from each content type---using a simulated streaming interface (see Figure~\ref{fig:screenshot}). Participants were compensated \$3 USD for completing the 5-minute study. The study protocol was reviewed and approved by our internal legal and ethics committee.

The interface implements a Parameter Estimation by Sequential Testing (PEST)~\cite{taylor1967pest} procedure to estimate each participant's natural reading speed. Participants were instructed to ``Read the text as it streams and try to understand it. When it pauses, tell us how the speed felt to you.'' At each pause, participants could indicate whether they preferred the speed to be faster or slower. Based on their response, the streaming continued, and the speed was adjusted by increasing (if chose ``prefer faster'') or reducing (if chose ``prefer slower'') $\Delta v$. The value of $\Delta v$ was also updated after each selection according to the standard PEST update rule.

\[
\Delta v =
\begin{cases}
\max\left(\Delta v_{\text{last}}/2,\ 0.2\right), & \text{if a choice reversal occurs} \\
\Delta v_{\text{last}}, & \text{if the choice is consistent}
\end{cases}
\]

The initial streaming speed was set randomly from 3 to 8 words per second (WPS), with an initial $\Delta v$ of 2 WPS.

After seven speed adjustments, participants were presented with an additional option: ``This is the same as my reading speed.'' They could either accept this speed or continue adjusting. The final selected speed was recorded as the participant’s comfortable reading speed for the corresponding passage.

In psycholinguistic research~\cite{marsden2018methodological}, similar self-paced reading methods are widely used to assess reading speed. In these tasks, participants read text on a computer screen and press a key to reveal each new segment, with the reading rate derived from their response times. However, because our goal is to more closely replicate the continuous streaming scenario typical of LLM interactions, we designed our system to stream a paragraph at a fixed speed and used the PEST procedure to determine each participant’s comfortable reading speed instead.

% This method is grounded in extensive psycholinguistic research that self-paced reading, where reading rate is derived from participants’ real-time responses, are widely used to assess reading speed and cognitive load~\cite{marsden2018methodological, jegerski2013self, jegerski2023using}, supporting our use of PEST-adjusted self-report as a valid proxy for natural reading speed.

To ensure data quality, we employed control checks to identify inattentive or inconsistent behavior. Participants were flagged for exclusion if they (a) consistently selected ``prefer faster'' regardless of the content, or (b) reported a faster preferred speed for the more complex (economics) passage than for the simpler (bedtime story) passage. The rationale for these criteria is twofold: (a) after seven consecutive ``faster'' selections, the speed would increase to approximately 20 WPS (1,200 WPM), which is well beyond the human reading limit even for the fastest readers; and (b) it is rare for someone to read a college-level text faster than a fifth-grade passage. In total, 21 out of 100 participants failed at least one of these control checks and were excluded from the analysis, which aligns with the 22.3\% failure rate reported in a previous study on crowd workers~\cite{saravanos2021hidden}.
%revision: clarify we did not discard fast reader

\paragraph{Results}

\begin{figure}[t]
  \centering
  \includegraphics[width=1.0\linewidth]{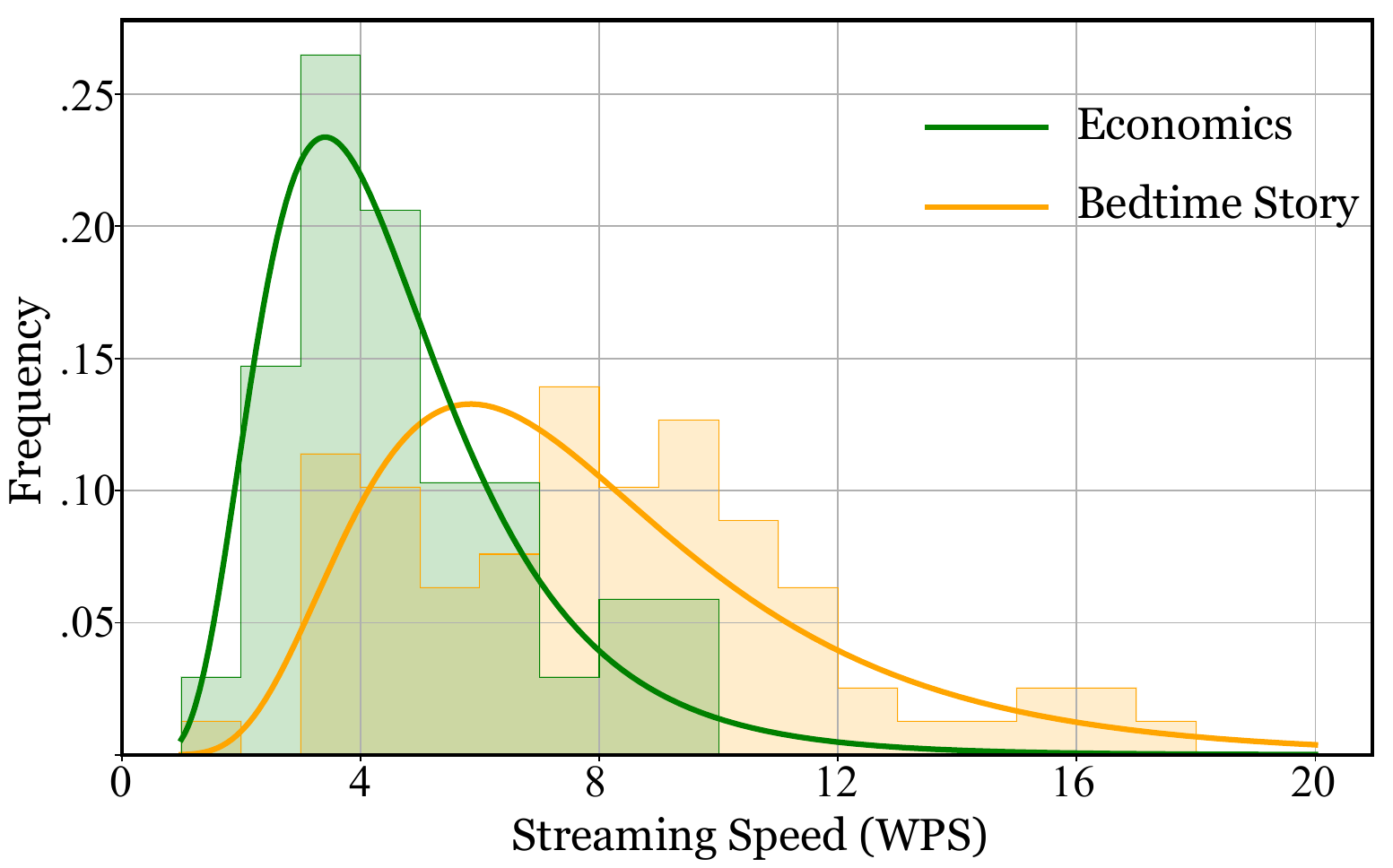}
  \caption{Histogram of preferred streaming speeds (in words per second) from the crowdsourced user study, overlaid with fitted log-normal distributions. The green curve represents the Economics content, while the orange curve corresponds to the Bedtime Story content.}
    \vspace{-3mm}
  \label{fig:crowdsource_distribution}
\end{figure}

We treat the final selected speed as the comfortable reading speed for each participant on each passage, and fit a log-normal distribution to the collection of these speeds.
Figure~\ref{fig:crowdsource_distribution} shows the histogram of the comfortable streaming speeds alongside the fitted log-normal curves. The green bars and curve correspond to the more complex economics passage, while the orange bars and curve represent the simpler bedtime story.
The peaks of both the histograms and the fitted distributions indicate that the comfortable reading speed for the complex passage is generally lower than that for the simpler one.

We further validated this difference using a paired t-test, which revealed a statistically significant result: $t(78) = 11.81$, $p < 0.001$. ($78$ is degrees of freedom correspond to $79$ effective paired samples.)
To evaluate the quality of the log-normal fit, we conducted a Kolmogorov-Smirnov (K-S) test~\cite{berger2014kolmogorov}. For both passages, the K-S statistic was 0.10, indicating a good alignment with the log-normal model. The corresponding $p$-values (0.35 and 0.39, respectively) suggest that the observed distributions do not significantly deviate from the log-normal assumption.

Using the log-normal distributions of reading speed inferred from the crowd-sourced data, we can compute the potential resource savings achieved through adaptive streaming for the two passages.
First, we estimate the savings in a scenario where computing resources are sufficient. We assume $s_{\text{max}}$ to be the average streaming speed of the GPT-4o API, which is approximately 45 WPS second~\cite{Haywood2024}.
The 99th percentile of the log-normal distribution is 21.20 WPS for the simpler passages and 11.97 WPS for the more complex one. By streaming at these speeds and using Equation~\ref{eq:saving} for computation, we can achieve a \textbf{63.14\%} reduction in computing resource usage while maintaining a 99\% SRAR.

Even with limited computing resources, similar to the scenario in Figure~\ref{fig:gaussian}(c), adaptive streaming can improve overall SRAR. The intersection of the two log-normal curves occurs at 5.53 WPS, suggesting that as long as the average computing budget is at least 5.53 WPS, adaptive streaming can lead to a net gain in user experience.

%% file: sections/3-system.tex
\section{System Implementation}

Section~\ref{sec:analysis} presented a statistical framework and experimental procedure that help understand the potential benefits of adaptive streaming. In that analysis, the distribution of the reading speed was obtained through an offline user study. But in a real-time LLM system, the content to be streamed is unknown until a user request is received. As a result, the corresponding reading speed distribution also cannot be determined in advance.

Prior studies have shown that human reading speed is largely influenced by the cognitive load imposed by the content~\cite{destefano2007cognitive, fink2001speed}. Building on this insight, a practical adaptive streaming system can use estimated cognitive load as a primary signal for determining appropriate streaming speed. To enable real-time adjustments, the system must estimate the cognitive load of each segment as it is being streamed and ensure that the streaming speed is allocated accordingly.

This section describes our system prototype, which includes a cognitive load estimation method and a resource allocation algorithm for adjusting streaming speed under budget constraints.

\subsection{Cognitive Load Estimation}\label{sec:cog_model}

Section~\ref{sec:cog_assess} discussed several cognitive load estimation methods for textual content, which span a wide range, from simple linguistic features that can be computed almost instantaneously, to more advanced machine learning models, or even hardware-based approaches. While hardware-based methods provide the most accurate measurement of cognitive load, they are not feasible for use in the client end of an cloud-based LLM service.
%Additionally, because cognitive load must be estimated in real time, we cannot rely on computationally expensive machine learning models.

In our prototype system, we explore two different cognitive load estimators. First, we use the Gunning-Fog Index~\cite{gunning1952technique}, a widely used readability metric that approximates the cognitive load of a text paragraph. Due to its simplicity, it can be computed with negligible overhead, making it suitable as a lightweight plugin for integration into LLM-based systems.

The second estimator leverages the LLM itself to assess cognitive load, inspired by the LLM’s strong performance on subjective, human-like evaluation tasks~\cite{gu2024survey}. Specifically, we provide the LLM with a definition of cognitive load based on Cognitive Load Theory and prompt it to append a cognitive load score using a special tag symbol, \texttt{<X>}, after generating a text segment. The number inside the brackets (e.g., \texttt{<3>}) represents the estimated cognitive load on a fixed scale, where lower values correspond to lower expected reading speeds, indicating more complex or cognitively demanding content. We constrain the score to a range from 1 to 10, where 1 denotes the highest cognitive load.

\vspace{3mm}
\textbf{Example of Generated Text:}
\begin{quote}
By introducing the concept of spacetime, the theory of relativity fundamentally transformed the landscape of modern physics.  \texttt{<3>} It unified space and time...
\end{quote}

During streaming, the system detects the tag symbol and extracts the inferred cognitive load score without displaying it to the user.
This method is particularly useful because it requires no additional external models and only adds the computation of a few tokens per segment. It can be seamlessly integrated into any existing LLM system by simply modifying the prompt instructions, making it a practical and scalable solution for real-time cognitive load estimation.

\subsection{Resource Allocation and Speed Modulation}\label{sec:alloc}

We formulate our resource allocation problem as follows. Suppose at a given time, the server is processing $n$ concurrent user requests, each corresponding to a distinct text segment to be streamed. Let the total available computational capacity at this moment be sufficient to stream $k$ words within a unit time interval. Without any speed modulation, each of the $n$ segments would be streamed uniformly at a speed of $k/n$ words per unit time.

To better align streaming speeds with estimated user cognitive load, we introduce adaptive speed modulation method. Specifically, let $s_i$ denote the cognitive load score for the $i$-th segment among all $n$ concurrent request. Our goal is to stream more cognitively demanding (lower-scoring) segments at slower speeds, providing users additional processing time, while allocating faster streaming speeds to segments with lower cognitive complexity (higher scores).

To achieve this, we first normalize the cognitive load scores into allocation weights ($s_i$) that sum to 1. We then apply a smoothed interpolation between a linear distribution of weights and a uniform distribution:

\begin{equation}\label{eq:alpha} w_i = \alpha \cdot \frac{s_i}{\sum_{j=1}^{n} s_j} + (1 - \alpha) \cdot \frac{1}{n}, \quad i = 1, 2, \dots, n \end{equation}

Here, $\alpha \in [0, 1]$ is a hyperparameter that controls the interpolation between a purely linear weighting (based on normalized cognitive load scores) and a uniform weighting (equal speed for all segments). When $\alpha = 1$, the weights are fully proportional to the cognitive load scores; when $\alpha = 0$, all segments receive equal weights regardless of content complexity.
The choice of $\alpha$ depends on the desired sensitivity of streaming speed to content complexity and can be determined empirically through a small-scale user study or by analyzing prior usage data. In our implementation, we set $\alpha=0.5$ for both the Gunning-Fog estimator and the LLM-based estimator, representing a balanced interpolation.

With these weights, the streaming speed $v_i$ (in words per unit time) allocated to each segment is computed as:

\begin{equation*} v_i = w_i \cdot k, \quad i = 1, 2, \dots, n \end{equation*}

This adaptive speed modulation algorithm ensures that computational resources are allocated in proportion to the cognitive complexity of each segment. The tunable parameter $\alpha$ provides both flexibility and interpretability to the resource allocation mechanism.

\subsection{Other Practical Considerations}

\paragraph{Backtracking Speed Control Token}
It is important to note that our current approach calculates the streaming speed allocation only after the server has already generated a given text segment. As a result, we cannot retroactively slow down the generation of that segment on the server side. A practical implementation of streaming speed control, therefore, is to let the client side stream the newly generated text at the determined speed, while the server pauses or adjusts the generation of subsequent sentences to align with the updated allocation. This ensures that users experience streaming at the calculated speeds in real time.
Implicitly, this approach assumes that adjacent sentences tend to have similar cognitive load. This idea connects to the notion of local coherence, a well-established concept in natural language processing that suggests neighboring sentences are systematically related in aspects such as sentence length, discourse structure, and lexical or topical continuity~\cite{schils1993characteristics, jurafsky2025chapter24}, which are factors that associated with the complexity of the textual content~\cite{schumacher2016predicting, davoodi2016contribution}.

% Empirical studies have also shown that these local coherence features can significantly improve text complexity predictions~\cite{schumacher2016predicting, davoodi2016contribution}.
%Implicitly, this approach assumes that within a text chunk, adjacent sentences will have similar cognitive load --- an assumption similar to that used in traditional readability calculations. 
%Readability formulas such as Flesch-Kincaid and Gunning Fog rely on the assumption that sentence-level features remain relatively consistent within a text sample, allowing average-based measures to provide stable estimates of reading difficulty~\cite{dale1948formula, gunning1952technique}.
%This assumption is also supported by empirical findings that surrounding context can significantly improve sentence-level readability prediction, suggesting that adjacent sentences tend to share similar difficulty and cognitive load~\cite{schumacher2016predicting}.

\paragraph{Modeling Computational Resource}
In our theoretical framework above, for simplicity, we treated generation speed as the primary computational resource to manage and allocate. However, in practical deployment, there are multiple ways to define and measure computational resources, including total cost budgets, available FLOPS~\cite{chen2023run}, the number of GPUs, or total power consumption. It is important to note that, depending on which resource model is used, the mechanism of speed modulation, and the relationship between computational efficiency and user experience can vary.

To best align our analytical model with practical implementation, one practical perspective is to treat power consumption as the resource budget being managed, with GPU clock frequency serving as the tunable parameter. In this approach, increasing the clock frequency results in faster model inference and higher power usage, while reducing the frequency slows down inference and conserves power. This technique, known as Dynamic Voltage and Frequency Scaling (DVFS)~\cite{le2010dynamic,  mei2017survey, tang2019impact}, is widely used in resource management for both CPUs and GPUs. On NVIDIA GPUs, power management can be achieved through the \texttt{nvidia-smi} API. Recent research also demonstrates a concrete example of dynamically tuning GPU frequency for efficient LLM serving~\cite{kakolyris2024slo}. By tuning the clock frequency, a service provider can adjust generation speed with a single control parameter, without requiring additional operations such as memory migration or task scheduling, which can incur extra overhead.

\paragraph{Speed Allocation Assumption}

Our speed allocation scheme (Equation~\ref{eq:alpha}) relies on one key underlying assumption: that under a fixed computational budget, streaming speed can be allocated linearly across different user requests. That is, decreasing streaming speed by $\Delta s$ for one request allows the provider to increase streaming speed by $\Delta s$ for another request. However, in real-world industry practice, the relationship between LLM inference speed and power consumption is complex and evolving, involving factors such as GPU types, LLM model architecture, and batch size. Given this complexity, modeling this phenomenon precisely is challenging without using simplifying assumptions.

Prior research suggests that GPU clock frequency is monotonically and positively correlated with LLM inference speed and power consumption~\cite{maliakel2025investigating, kakolyris2024slo}. This positive correlation still supports the validity of our method: reducing the speed for more complex text can free up resources to increase speed for simpler text, thereby improving SRAR, even if the trade-off may not be exactly one-to-one. Additionally, one recent investigation~\cite{maliakel2025investigating} and an industry report from the NVIDIA GTC 2025 Keynote~\cite{wu2025testtimecompute} found that within certain ranges, power consumption may scale linearly with LLM inference speed. We emphasize that detailed modeling of these effects is complex and beyond the scope of this study. For tractability, we adopt the simplified linear allocation assumption throughout all our analyses and simulations, with the goal of providing theoretical insights that can inform the practical design of our adaptive streaming method.

%% file: sections/4-evaluation.tex
\section{Crowdsourcing Experiment} \label{sec:crowdsourcing}

In this study, we aim to address two central research questions regarding our proposed adaptive streaming system: First, how effectively does our cognitive load estimation method perform under realistic usage scenarios? Second, how does our adaptive streaming approach compare to the baseline method of streaming content at a uniform speed, in terms of computational resource efficiency and SRAR?

% We adopt the following methodology to evaluate our system. Consider a scenario where an LLM simultaneously streams $N$ distinct text segments, and each segment's comfortable reading speed follows an independent log-normal distribution based on our previous analysis. We can apply the user study protocol established in Section~\ref{sec:real} to empirically obtain the reading speed distributions for all $N$ segments. 

To investigate these questions, our system combines cognitive load estimation with the adaptive speed modulation algorithm described in Section~\ref{sec:alloc} to determine appropriate streaming speeds for different text passages under varying computational resource constraints.
To provide empirical grounding, we conducted a crowdsourcing study to obtain distributions of users' comfortable reading speeds of the passages. Once these distributions are collected and streaming speeds assigned, we compute the overall SRAR---defined as the proportion of users who receive streaming speeds equal to or greater than their natural reading speeds (recall the definition at Section~\ref{sec:analysis})---at each level of computational budget, for all speed modulation methods.
Finally, we simulate and compare the computational resources required by each method to achieve equivalent SRAR.
%remove satisfaction

% As a comparative baseline, we also stream all content uniformly at a single speed under the same computational budget. 

\subsection{Data Collection}
The crowdsource user study conducted here follows a procedure similar to the one described previously in Section~\ref{sec:real}. To construct a diverse dataset, we first prompted GPT-4o to generate 10 English text passages spanning various randomly selected topics. We also explicitly instructed the LLM to produce content of varying cognitive loads across passages. For calibration, the simplest content is set as a bedtime story suitable for a six-year-old child, while the most complex content is an explanatory passage about a philosophical concept. All other topics and corresponding passages were generated autonomously by the LLM within this defined complexity range. To ensure quality, we verified that each passage could be understood without requiring specific technical knowledge, with lengths ranging consistently between 150 and 200 words.

We recruited 200 native English speakers through the Prolific platform. Participants consisted of 44\% male and 56\% female, with an average age of 38.6 years (ranging from 24 to 74). Among participants, 57\% reported holding a bachelor's degree or higher.  During the study, each participant was presented with 6 randomly selected passages using the same simulated streaming interface and PEST-based speed adjustment procedure described in Section~\ref{sec:real}. The passage order was fully randomized across participants to avoid order effects. Each session lasted about 15 minutes to reduce fatigue and minimize fluctuations in reading speed, and each participant was paid \$7 USD for the study.

We applied similar control checks as described in Section~\ref{sec:real} to ensure data quality. Specifically, participants were flagged if they selected “faster” for every adjustment within a passage or if their preferred reading speed for the most complex passage was faster than for the easiest passage. In addition to these checks, after finishing each passage, the passage would disappear, and we then asked a two-choice comprehension question, generated by the LLM, to verify whether participants understood the content. For example, one passage tells a short story about a child baking cookies with their grandmother, and the corresponding question is: ``What food is the child trying to make?'' All control checks needed to be satisfied for a data point to be considered valid.
After applying these quality checks, we ensured that each passage received between 80 and 100 valid data points representing individual users’ comfortable streaming speeds.

% \paragraph{Hyper-parameter Estimation}

\begin{figure}[t]
  \centering
  \includegraphics[width=1.0\linewidth]{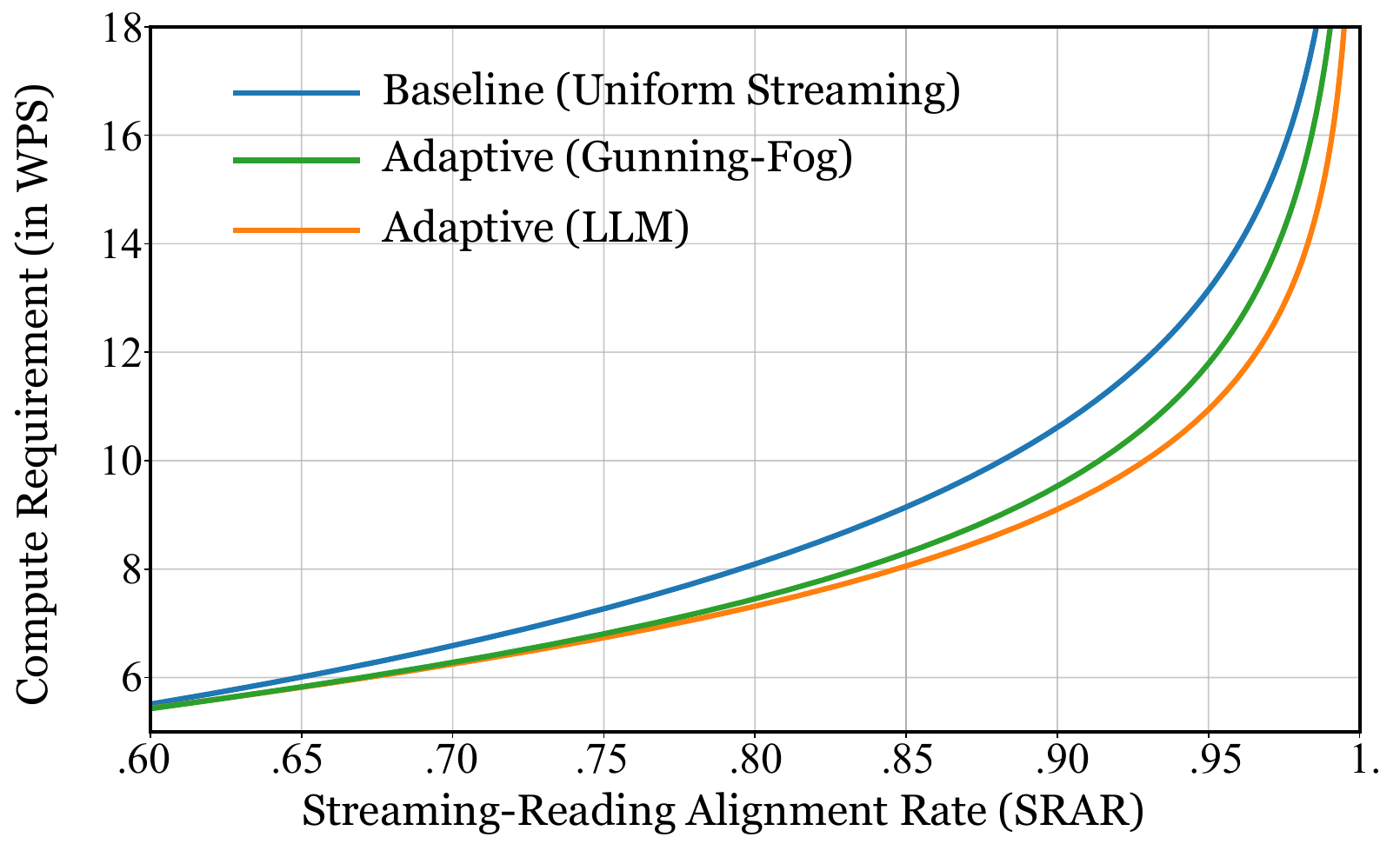}
  \caption{Compute requirements of different methods at varying SRAR. The compute requirements are measured by average streaming speed in words per second (WPS), where lower is better. At the same overall SRAR, our proposed methods (Gunning-Fog and LLM) consistently show lower compute requirements compared to the baseline.}
    \vspace{-3mm}
  \label{fig:speed_satisfaction}
\end{figure}

% As introduced in Section~\ref{sec:alloc}, our adaptive speed modulation method requires tuning a hyperparameter $\alpha$, which controls the balance between resource allocation based on estimated cognitive load scores and a uniform distribution. 

% For a given computational budget $k$, the baseline method is a uniform allocation in which every segment is streamed at a constant speed of $k$ WPS. Using the fitted log-normal reading speed distribution for each passage, we compute the average user satisfaction rate achieved by this baseline across all 10 passages.

% Similarly, for each cognitive load estimation method under each computational budget $k$, we compute the average user satisfaction rate as a function of $\alpha$. We sweep $\alpha$ from 0 to 1 in increments of 0.01 to determine the value that yields the highest satisfaction rate.

% When using an LLM-based cognitive load estimator to guide resource allocation, we found that the optimal value of $\alpha$ is 0.52. This setting results in a 6.52\% relative improvement in user satisfaction over the baseline at a computational budget of 8 WPS (baseline: 79.49\%, LLM-based modulation: 84.67\%).
% Using the Gunning-Fog Index as the cognitive load estimator, the optimal $\alpha$ is 0.44, yielding a 5.00\% relative improvement at a computational budget of 7.6 WPS (baseline: 77.14\%, Gunning-Fog-based modulation: 80.99\%).

% These results suggest that an $\alpha$ value around 0.5—representing a balanced interpolation between uniform allocation and strict cognitive score-based allocation—achieves the best modulation performance.

\subsection{Results and Analysis}

\paragraph{Validation of Cognitive Load Estimation Methods}
To first evaluate the quality of our cognitive load estimation approaches, we examined their correlation with empirically measured comfortable reading speeds. Specifically, we calculated Pearson's correlation coefficients between each method’s cognitive load scores and the median comfortable reading speeds obtained from our crowdsourcing experiment for each passage. The results demonstrated strong and statistically significant correlations for both methods: the LLM-based cognitive load estimation achieved a correlation of ( r = 0.955, p < 0.001 ), whereas the Gunning-Fog Index yielded a correlation of ( r = 0.828, p = 0.003 ). These results confirm that both cognitive load estimation methods effectively approximate users' comfortable reading speeds, with the LLM-based approach exhibiting notably higher predictive accuracy.

\paragraph{Comparison of Resource Efficiency and SRAR}

To better illustrate the overall effectiveness of cognitive-aware resource allocation, we plotted the resource utilization of different methods across varying SRAR levels in Figure~\ref{fig:speed_satisfaction}. The x-axis represents the SRAR, while the y-axis indicates computational resource usage, defined as the average streaming speed (in WPS) across all concurrently streamed passages required to achieve each SRAR level. Lower values on the y-axis indicate greater resource efficiency.

As shown in Figure~\ref{fig:speed_satisfaction}, at equivalent SRAR, the Gunning-Fog-based allocation method (green line) consistently requires fewer computational resources compared to the uniform-speed baseline (blue line). Moreover, the LLM-based allocation method (orange line) yields even greater efficiency gains. Notably, at a high SRAR target of 95\%, the Gunning-Fog-based method achieves a 10.33\% reduction in computing resources relative to the baseline, while the LLM-based adaptive modulation attains an even more substantial reduction of 16.79\%. Table~\ref{tab:compute_reduction} presents the exact average streaming speed required by each method to achieve the target SRAR, as well as the compute savings ratio of our proposed methods compared to uniform streaming.

These findings highlight clear efficiency improvements achieved through adaptive streaming. The Gunning-Fog Index, as a straightforward off-the-shelf metric, incurs negligible computational cost while still providing a meaningful increase in streaming efficiency. On the other hand, the LLM-based estimation method, which produces cognitive load scores more closely aligned with actual human reading speeds, achieves even greater overall efficiency improvements.

% Thus, the choice between these two cognitive load estimation methods ultimately depends on the specific constraints and requirements of the deployment environment. In scenarios where minimizing computational overhead is critical, the Gunning-Fog Index offers a practical and effective solution. Conversely, when computational resources are more abundant or the focus is maximizing efficiency gains, the LLM-based cognitive load estimation is the recommended choice.

% [optimizing in the low satisfactions region should be avoided, as too slow would even impact user experience larger]
We also observe that when the streaming speed is lower than $6$ WPS, there is little difference between uniform streaming and adaptive streaming. This corresponds to the scenario illustrated in Figure~\ref{fig:gaussian}(c), where adaptive streaming may result in more users experiencing delays. The value of $6$ WPS provides an empirical insight from real data into the threshold at which the benefits of adaptive streaming begin to diminish.

\begin{table}[t]  
\centering  
\small  
\begin{tabular}{c|c|cc|cc}
\toprule
\multirow{2}{*}{\textbf{\shortstack{SRAR}}} & \textbf{Baseline} & \multicolumn{2}{c|}{\textbf{Gunning-Fog}} & \multicolumn{2}{c}{\textbf{LLM}} \\
\cmidrule(r){2-2} \cmidrule(r){3-4} \cmidrule(r){5-6}
 & \textbf{Compute} & \textbf{Compute} & \textbf{Save \%} & \textbf{Compute} & \textbf{Save \%} \\
\midrule
0.65 & 6.01 & 5.83 & 3.00\% & 5.82 & 3.16\% \\  
0.70 & 6.59 & 6.28 & 4.70\% & 6.25 & 5.16\% \\  
0.75 & 7.27 & 6.81 & 6.33\% & 6.74 & 7.29\% \\  
0.80 & 8.10 & 7.46 & 7.90\% & 7.32 & 9.63\% \\  
0.85 & 9.15 & 8.30 & 9.29\% & 8.06 & 11.91\% \\  
0.90 & 10.62 & 9.54 & 10.17\% & 9.11 & 14.22\% \\  
0.95 & 13.16 & 11.80 & 10.33\% & 10.95 & 16.79\% \\  
\bottomrule  
\end{tabular}    
\caption{Reduction in computational resource usage (\%) achieved by cognitive load estimators compared to the uniform streaming baseline at varying SRAR. ``Compute'' is measured as average streaming speed in WPS, with lower values indicating lower resource usage.}  
\label{tab:compute_reduction}  
\end{table}  

%% file: sections/5-conclusion.tex
%!TEX root = proceedings.tex

%%
%% NOTE: Assign collaboration badges and section labels to all sections and 
%% subsects when created (badges include: \incomplete, \underRevision,
%% \readyForFeedback, \feedbackProvided, \complete, \locked)
%%
\section{Applications, Limitations, and Future Directions}
\label{sec:discussion}
While our framework demonstrates promising results, there remain several limitations and opportunities for further development across application scenarios, model design, and system-level optimization.

\paragraph{Application Scenario}
In principle, our adaptive streaming framework requires providers to deploy LLMs on their own GPU infrastructure to control generation speed. Such providers includes major services such as ChatGPT, Google Gemini, Claude, and DeepSeek, as well as any organization deploying open-source LLMs on their own infrastructure. Nowadays, many services developed by startups or individual developers are powered by open-source LLMs that are fine-tuned for specific conversational AI scenarios~\cite{VentureBeatOpenSourceLLMs2024}, such as customer support, educational chatbots, or personal assistants. Unlike large companies, these providers often have more limited GPU resources. Employing our method could help them make better use of their resources and enable higher service bandwidth.

We also acknowledge that certain LLM use cases may not benefit directly from adaptive streaming based on cognitive load. Specifically, tasks that utilize LLM outputs primarily as intermediate computational steps, such as code generation, data processing pipelines, or internal reasoning procedures, typically prioritize rapid delivery with users focusing primarily on the final results rather than incremental outputs. In these contexts, users prefer the model to stream tokens at the maximum available speed, making cognitive-aware streaming less relevant.

Nevertheless, our adaptive streaming mechanism remains highly applicable to general-purpose conversational AI through the integration of an intention detection module following the user's prompt. For instance, when users request explanatory content (``Explain the theory of relativity''), instructional guidance (``How to make pasta dough''), or in-depth informational answers ("What are the symptoms and treatments for diabetes?"), the intention detector identifies these interactions as reading-intensive tasks suitable for adaptive streaming. Conversely, task-oriented interactions (e.g., ``Generate Python code for sorting algorithms'') bypass adaptive streaming, allowing for immediate, maximum-speed token delivery and efficient resource allocation.

Moreover, there has been extensive recent research and development of LLM-enabled interactive conversational systems designed to assist users through detailed textual interactions. Representative applications include AI-powered educational tutors delivering personalized learning content~\cite{zhu2025autopbl}, virtual healthcare assistants providing medical information and instructions~\cite{ramjee2025ashabot}, emotional support chatbots~\cite{zheng2025customizing}, and physical activity coaching systems~\cite{jorke2025gptcoach}. In these scenarios, users typically engage deeply with the content and often ask follow-up questions. Deploying our cognitive-aware adaptive streaming framework in such applications could help these systems serve more users simultaneously while maintaining a high-quality user experience. By aligning streaming rates with users’ processing capabilities, our approach directly contributes to ongoing efforts in HCI to design more responsive, scalable, and resource-efficient conversational AI services.
%add more HCI applications and citations

% Moreover, our proposed adaptive streaming system broadly benefits conversational AI applications designed to assist users through detailed textual interactions. Representative scenarios include AI-powered educational tutors delivering personalized learning content, virtual healthcare assistants providing medical information and instructions, and customer support chatbots responding to detailed product-related inquiries. In such applications, users typically engage actively with the content and may interactively ask follow-up questions. Deploying our cognitive-aware adaptive streaming framework in these contexts enhances both user experience and system efficiency, by aligning streaming rates with users' processing capabilities.

\paragraph{Cognitive Load/Reading Speed Prediction}
As a prototype implementation, we currently use linguistic features and the LLM itself to predict cognitive load. 
While this approach demonstrates some effectiveness, it may not precisely model the user’s cognitive load on every aspect.  
One direction for future work is to develop a model that can make fast predictions while also estimating the distribution of reading speeds more accurately, thereby supporting more efficient resource allocation.

Another promising avenue is adapting the system to specific application scenarios. For instance, if the LLM is deployed to provide information about a particular product to customers, the service could analyze users' historical requests alongside the corresponding LLM-generated responses. This data could then be leveraged to model the distribution of users' reading speeds. Given the specific focus of such applications, the estimation could be accurately performed with a small-scale user study. The resulting distribution would be useful in optimizing resource allocation for that application in the future.

Additionally, emerging AR/VR headsets typically include gaze-tracking capabilities, which could be used to directly measure users' reading speeds. For AI assistant applications deployed on AR/VR headsets, our adaptive streaming technique could leverage gaze tracking data to dynamically adjust the streaming rate, ensuring that the content delivery closely matches the user's consumption speed.

\paragraph{Resource Allocation with Low-level Mechanism}
In our current study, we treat computational resource allocation as a speed bandwidth optimization problem. This can be practically implemented using DVFS, which enables us to evaluate our adaptive streaming method in a controlled environment and analyze its performance. In real-world deployments, resource optimization may also involve additional factors related to low-level system mechanisms such as data transmission, hardware constraints, and cache management. In future work, it would be valuable to incorporate these factors and explore joint optimization strategies and integrate our method with existing serving libraries such as vLLM~\cite{kwon2023efficient}, to further enhance streaming efficiency.

\paragraph{Exploring Different Speed Adjusting Mechanism}
Currently, our method uses speed modulation to align with users’ reading speeds. However, users may engage with content in different ways. For example, after reading logically complex content such as mathematical formulas, users may pause and reflect before continuing. Inspired by this observation, it would be interesting to explore alternative streaming strategies, such as streaming at full speed followed by brief pauses, or combining variable speed modulation with pause mechanisms. Investigating how different streaming strategies affect user experience represents a compelling direction for future research.

\paragraph{Diversifying User Experience Metrics}
Our current framework models the optimization problem with the goal of maintaining an overall SRAR, which is based primarily on statistical modeling and simulations. However, broader measures of user experience, such as perceived comfort, engagement, subjective workload ratings, and qualitative feedback, could offer richer insights. For example, our current approach does not explicitly factor in individual users' waiting times; it only considers whether the streaming speed falls below a user's comfortable threshold, without considering how far below that threshold the speed actually is. In the future, we could incorporate a weighted metric based on the magnitude of the difference between the actual streaming speed and the user's comfortable speed. Optimizing the system using such refined metrics could help achieve a more fine-grained balance between resource efficiency and user experience.

\paragraph{Distribution of Readers}
Our current analytical framework assumes that reading speeds within a user group follow a log-normal distribution. This assumption has been empirically validated by our experiment using a K-S test, as described in Section ~\ref{sec:real}. However, modeling human reading speed for any specific population remains a complex challenge, as it must account for diverse conditions such as dyslexia~\cite{shaywitz2005dyslexia} or unusually fast readers~\cite{rayner2016so}, who may process information in fundamentally different ways. The characteristics of the user population can also vary significantly depending on the specific application scenario. In this paper, our goal is to provide a technical framework for implementing such a system, along with an example approach for evaluating its effectiveness, rather than offering a comprehensive study of human reading behavior. We hope this framework serves as a practical example that practitioners can adapt to analyze performance in their own settings.
%discuss we assume a log-normal distribution, but there are more potential possible distribution, and it is still hard to obtain the concrete form of distribution, and many research has many different conclusion of the exact form of distribution

% \subsection{Limitations and Future Work}

% \textcolor{red}{Potential future direction: 1. domain specific scenario for better cognitive load prediction. 2. better fast reading speed prediction model. 3. better schedule mechanism that integrated with low level. 4. persona aware streaming (non native speaker, knowledge level, education level) 5. only useful for readable text streaming, not for reasoning or code generation. In some task, user may want the results generated as soon as possible, but also pose new oppotunity to detect intent (what part need read, what part don't 6. integrating cognitive detection mechanism in post-train, so no need extra cognitive detection model). 7. different modulation mechanism, such as pausing.}

% \textcolor{red}{application: online tutorial}

\section{Conclusion}

To summarize, we propose the first method that leverages the cognitive load imposed on users to optimize the efficiency of LLM serving. While LLMs can stream tokens at high speeds, the output is not instantaneous---making the streaming rate a non-negligible factor in user interaction. Our adaptive streaming approach uses cognitive load signals to dynamically adjust token delivery speed, improving computational efficiency without largely impacting the user experience. Our crowdsourced user study demonstrates that this method increases the proportion of users whose streaming experience aligns with their content-dependent reading pace.

We hope this work encourages further exploration of user-centered and resource-aware LLM serving strategies, and inspires future research at the intersection of cognitive modeling, system optimization, and interactive AI design.